\documentclass[12pt]{iopart}

\usepackage{iopams,amstext}
\usepackage[dvips]{graphicx}

\newcommand{\quoting}[1]{``#1''}
\renewcommand*{\vec}[1]{\mathbf{#1}}

\begin{document}

%%%%%%%%%%%%%%%%%%%%%%%%%%%%%%%%%%%%%%%%%%%%%%%%%%%%%%%%%%%%%%%
% title, authors
%%%%%%%%%%%%%%%%%%%%%%%%%%%%%%%%%%%%%%%%%%%%%%%%%%%%%%%%%%%%%%%

\title{Measurement of the Goos-H{\"a}nchen shift in a microwave cavity}

\author{J Unterhinninghofen$^1$, U Kuhl$^{2,3}$, J Wiersig$^1$, H-J St{\"o}ckmann$^2$, and M Hentschel$^4$}
\address{$^1$ Institut f{\"u}r Theoretische Physik, Otto-von-Guericke-Universit{\"a}t Magdeburg, Postfach 4120, D-39106 Magdeburg, Germany}
\address{$^2$ Fachbereich Physik, Philipps-Universit{\"a}t Marburg, Renthof 5, D-35032 Marburg, Germany}
\address{$^3$ Laboratoire de Physique Mati{\`{e}}re Condens{\'{e}}e (LPMC/CNRS UMR6622), Universit{\'{e}} Nice-Sophia Antipolis, Parc Valrose, F-06108 Nice Cedex 2, France}
\address{$^4$ Max-Planck-Institut f{\"u}r Physik komplexer Systeme, N{\"o}thnitzer Str. 38, D-01187 Dresden, Germany}
\ead{julia.unterhinninghofen@ovgu.de}

\begin{abstract}
	We present measurements of the Goos-H{\"a}nchen shift in a two-dimensional dielectric microwave cavity. Microwave beams are generated by a suitable 		superposition of the spherical waves generated by an array of antennas; the resulting beams are then reflected at a planar interface. By measuring the 		electric field including its phase, Poynting vectors of the incoming and reflected beams can be extracted, which in turn are used to find the incoming angle 		and  the positions where the beam hits the interface and where it is reflected. These positions directly yield the Goos-H{\"a}nchen shift. The results are 		compared to the classical Artmann result  and a numerical calculation using Gaussian beams.
\end{abstract}


\submitto{\NJP}

%%%%%%%%%%%%%%%%%%%%%%%%%%%%%%%%%%%%%%%%%%%%%%%%%%%%%%%%%%%%%%%%%%%%%

%\maketitle

%%%%%%%%%%%%%%%%%%%%%%%%%%%%%%%%%%%%%%%%%%%%%%%%%%%%%%%%%%%%%%%%%%%%%%
% Paper text
%%%%%%%%%%%%%%%%%%%%%%%%%%%%%%%%%%%%%%%%%%%%%%%%%%%%%%%%%%%%%%%%%%%%%%

\section{Introduction}
\label{sec:introduction}

The Goos-H{\"a}nchen shift (GHS), the lateral shift of a totally reflected light beam at a dielectric interface, was discovered by Goos and H{\"a}nchen in 1947 \cite{Goos1947}. It has subsequently been observed not only in optics, but also in acoustics \cite{Goos1947,Lai1986} and neutron beams \cite{Haan2010}. The quantum analog has been observed at graphene interfaces \cite{Beenakker2009}. A simple theoretical model is due to Artmann \cite{Artmann1948}; there, the shift is calculated analytically for a beam consisting of two plane waves with slightly different propagation directions. Analytical calculations of the GHS for Gaussian beams have also been performed \cite{Lai1986,Artmann1951} as well as numerical ones \cite{Aiello2008}.

The GHS is difficult to measure directly for optical wavelengths because the shift is on the order of the wavelength and thus very small. Goos and H{\"a}nchen used multiple reflections; recently, a measurement technique \cite{Schwefel2008} has been proposed which only needs one reflection and measures the GHS at a dielectric interface relative to a metallic one with a CCD camera. 

Because the GHS scales with the wavelength $\lambda$ of the beam, and thus vanishes in the limit $\lambda\rightarrow 0$, it can be seen as a first-order wave correction to geometrical (ray) optics. Such corrections have recently been incorporated into a ray model for optical microresonators \cite{Hentschel2002,Schomerus2006,Unterhinninghofen2008,Altmann2008,Unterhinninghofen2010} in order to explain deviations of calculations of optical modes from ray predictions. The resulting \quoting{extended ray models} are interesting tools which can help to understand ray-wave correspondence in optical microcavities. Moreover, as microcavities can be viewed as open quantum billiards \cite{Noeckel1997}, this is directly related to the study of quantum-classical correspondence in open quantum systems \cite{Stoeckmann2000}.

Unfortunately, there are no experiments which directly measure the GHS in optical microcavities, as the measurement of modes with high spatial resolution is difficult in these systems. Moreover, all calculations have assumed reflection at planar interfaces or circular interfaces, even though the interfaces in microcavities typically have non-constant curvature.

High-resolution measurements of electric fields, including the phase of the field, can be done easily in \emph{microwave} resonators \cite{Stein1992,Stein1995,Kuhl2007}; they thus could become model systems where predictions for the influence of the GHS on modes in optical microcavities could be tested. It is also possible to do microwave GHS measurements at curved interfaces and directly study deviations from the results for planar or circular interfaces.
The influence of boundary curvature is of particular interest in optical microcavities, as cavities with dimensions on the order of the wavelength of the optical modes have recently been fabricated \cite{Song2009}; in such systems, the boundary curvature is also on the order of the wavelength and typically not negligible.

In this paper, we demonstrate that such measurements are in principle possible by considering the simplest scenario. We prepare microwave beams with a well-defined propagation direction and spatial extension, and measure their GHS upon reflection at a planar interface, comparing the results to both the well-known Artmann result and a Gaussian beam calculation.

The paper is structured as follows. The experimental setup is described in \sref{sec:experimental_setup}, and the generation of suitable microwave beams is described in \sref{sec:beam_generation}. The resulting beams and the GHS extracted from them are shown in \sref{sec:results} and compared to theoretical predictions.

%%%%%%%%%%%%%%%%%%%%%%%%%%%%%%%%%%%%%%%%%%%%%%%%%%%%%%%%%%%%%%%%%%%%%%
% Setup
%%%%%%%%%%%%%%%%%%%%%%%%%%%%%%%%%%%%%%%%%%%%%%%%%%%%%%%%%%%%%%%%%%%%%%

\section{Experimental setup}
\label{sec:experimental_setup}

\begin{figure}
		\includegraphics[width=0.8\textwidth]{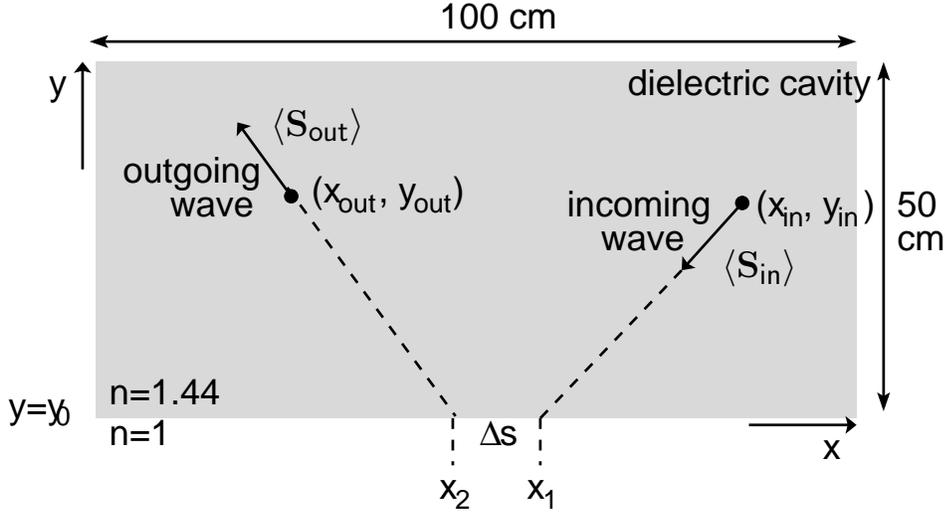}
		\caption{\label{fig:ghs_extraction} Extraction of the GHS~$\Delta s$ from the measured incoming and outgoing beams. 
		$\langle\vec{S}_{\text{in}}\rangle$ and $\langle\vec{S}_{\text{out}}\rangle$ are averaged Poynting vectors of the incoming and outgoing beams and
		$(x_{\text{in}},y_{\text{in}})$ and $(x_{\text{out}},y_{\text{out}})$ position averages. $x_1$ and $x_2$ are intersections of the straight lines 			defined by the Poynting vectors and averaged positions with $y=y_0$.}
	\end{figure}

	Experiments are done using a rectangular (50~cm~$\times$~100~cm) cavity made of Teflon with a refractive index $n=1.44$. The vertical dimension of the 
	cavity is $h=0.5$~cm; modes with frequencies below $\nu_{\text{max}}=c/(2nh)\approx 20.8$~GHz with the vacuum speed of light $c$ can thus be regarded as 
	two-dimensional. The electric field is measured using an Agilent 8720ES vector network analyzer. The generated beam travels to one boundary of the cavity 		and are reflected; the GHS~$\Delta s$ can directly be extracted as the shift between incoming and outgoing (reflected) beams.
	Figure~\ref{fig:ghs_extraction} shows the idea of GHS extraction from measured data. It is possible to extract Poynting vectors 
	$\vec{S}_{\text{in}}$ and $\vec{S}_{\text{out}}$ of the incoming and outgoing beams at each point~$\vec{r}=(x,y)$ in space from the measured 
	electric fields. We define the propagation direction of the incoming and outgoing beams as the average Poynting vectors~$\langle\vec{S}_{\text{in}}\rangle$, 		$\langle\vec{S}_{\text{out}}\rangle$.
	Because our measurements are done on a quadratic grid, they are given by
	\begin{equation}
	\label{eqn:average_poynting_def}
		\langle\vec{S}_{\text{in (out)}}\rangle=\frac{1}{N_x N_y}\sum_{i=1}^{N_x}\sum_{j=1}^{N_y} \vec{S}_{\text{in (out)}}(x_i,y_j),
	\end{equation}
	where the $x_i$, $y_j$ are the positions on which $\vec{S}_{\text{in (out)}}$ values have been measured. $N_{x (y)}$ is the number of
	$x (y)$ positions measured.
		
	The average Poynting vectors define the propagation direction of the incoming and outgoing beams. Together with the position 
	$\vec{r}_{\text{in}}=(x_{\text{in}},y_{\text{in}})$ on the
	incoming beam ($\vec{r}_{\text{in}}$ is the average over all positions at which the incoming beam is measured), $\langle\vec{S}_{\text{in}}\rangle$
	defines a straight line
	\begin{equation}
	\label{eqn:straight_line_S}
		\pmatrix{%
			x \cr y
			}
		=\pmatrix{%
					x_{\text{in}} \cr
					y_{\text{in}}
			 } +t\,\langle\vec{S}_{\text{in}}\rangle
	\end{equation}
	parametrised by $t$.
	The intersection of this straight line with the bottom of the Teflon plate $y=y_0$ yields the position~$x_1$ where the incoming beam hits the cavity 		boundary. Analogously, the 
	position~$x_2$ where the outgoing beam starts at the boundary can be calculated by intersecting the straight line
	\begin{equation}
	\label{eqn:straight_line_S2}
		\pmatrix{%
			x \cr y
		}=\pmatrix{%
					x_{\text{out}} \cr
					y_{\text{out}}
				} +t\,\langle\vec{S}_{\text{out}}\rangle
	\end{equation}
	with~$y=y_0$. The GHS is then given by $\Delta s=x_2-x_1$.

%%%%%%%%%%%%%%%%%%%%%%%%%%%%%%%%%%%%%%%%%%%%%%%%%%%%%%%%%%%%%%%%%%%%%%
% beam generation
%%%%%%%%%%%%%%%%%%%%%%%%%%%%%%%%%%%%%%%%%%%%%%%%%%%%%%%%%%%%%%%%%%%%%%

\section{Beam generation}
\label{sec:beam_generation}

	Microwave antennas produce spherical waves; for the GHS measurements, beams approximating plane waves have to be generated from these spherical waves. The generation of \emph{plane waves} can be easily done
	because the superposition of $N$ spherical waves $\psi_j$ with wave number $k$ and centers $(x_j,y_j)$ on a straight line creates a beam
	\begin{equation}
	\label{eqn:plane_wave_many_ant}
		\psi(x,y)=\sum_{j=-N/2}^{j=N/2}\psi_j(x,y)=\sum_{j=-N/2}^{j=N/2}\frac{\exp\left[ ik\sqrt{(x-x_j)^2+(y-y_j)^2}\right]}{\sqrt{(x-x_j)^2+(y-y_j)^2}}
	\end{equation}
	which approaches a plane wave if $N$ goes to infinity. A similar generation of a suitable beam from multiple spherical waves generated by microwave antennas 		has been used in \cite{OriaIriarte2010}.

	The propagation direction of a plane wave generated from spherical waves can be influenced by adding a phase factor $\phi(j)$ to each spherical wave. In 		this case,
	the generated beam is given by
	\begin{equation}
	\label{eqn:plane_wave_phase_factors}
		\psi(x,y)=\sum_{j=-N/2}^{j=N/2} \psi_j(x,y)e^{i\phi(j)}
	\end{equation}
	By varying $\phi(j)$, one can achieve propagation directions which lead to reflection under different angles of incidence. I.e., $\phi(j)=j$
	leads to reflection with an angle below the critical angle $\chi_{\text{cr}}\approx 44$~degrees for Teflon with $n=1.44$, $\phi(j)=-j$ leads to reflection
	with an angle above the critical angle, and for $\phi(j)=0$ the reflection happens close to the critical angle. Unfortunately, there is no analytical
	formula for the relation of $\phi(j)$ and the resulting propagation direction; the choice of the $\phi(j)$ functions therefore remains somewhat arbitrary.

	A plane wave, however, does not experience the GHS upon reflection, as the GH effect is a consequence of the \emph{interference} of partial waves
	with different angles of incidence. Creating a plane wave thus does not suffice if one wants to measure the GHS. But by superposing \emph{two}
	plane waves generated according to \eref{eqn:plane_wave_phase_factors} with different $\phi(j)$ functions corresponding to a small difference
	in their propagation directions leads to a beam just like the one assumed in the derivation of the Artmann result:
	\begin{equation}
	\label{eqn:exp_artmann}
		\psi(x,y)=\psi_1(x,y)+\psi_2(x,y)\approx\exp\left(ikp_1x\right)+\exp\left(ikp_2x\right),
	\end{equation}
	with $p_m=\sin\chi_m\sim S_{m,x}$, where $S_{m,x}$ is the $x$ component of the Poynting  vector of partial wave $m$ (with $m=1,\,2$).

	As both partial waves of the beam~\eref{eqn:exp_artmann} are calculated using the same spherical wave components~$\psi_j$,
	\begin{equation}
	\label{eqn:exp_artmann2}
		\psi_1=\sum_{j=-N/2}^{N/2} \psi_j e^{i\phi_1(j)}, \, \psi_2=\sum_{j=-N/2}^{N/2} \psi_j e^{i\phi_2(j)},
	\end{equation}
	the sum is just given by
	\begin{equation}
	\label{eqn:exp_artmann3}
		\psi=\psi_1+\psi_2=\sum_{j=-N/2}^{N/2} \psi_j \left(e^{i\phi_1(j)}+e^{i\phi_2(j)}\right).
	\end{equation}

\begin{figure}
		\includegraphics[width=\textwidth]{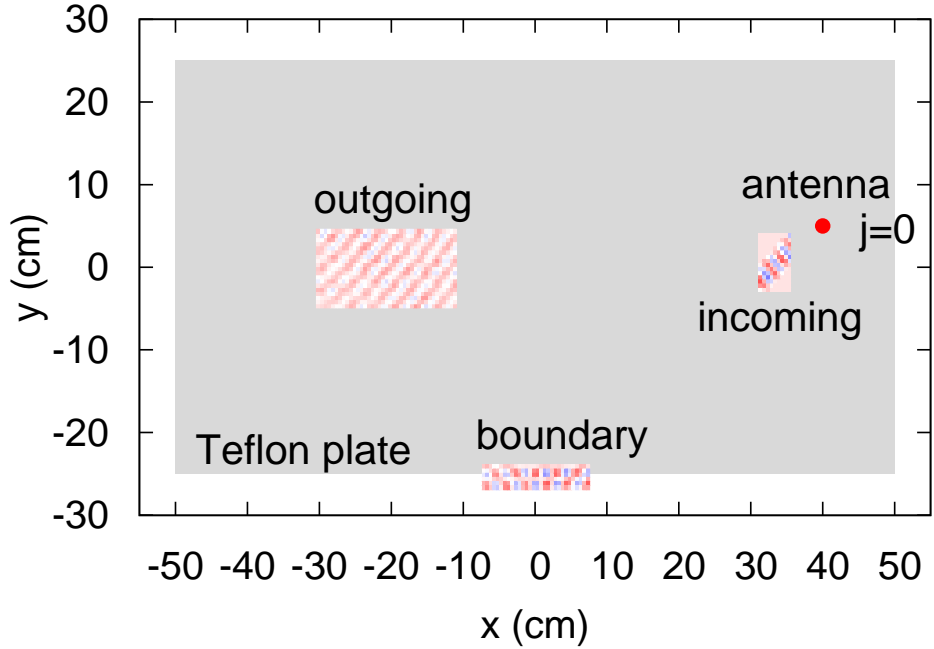}
		\caption{\label{fig:whole_billiard_wave} Measured real parts $\text{Re}(\psi)$ of the beams at $\nu=15$~GHz. The plate is indicated as the gray 		region; the position $(x_A,y_A)$ of the middle ($j=0$) antenna in the antenna array is shown as well.}
	\end{figure}

	In the experiment, measurements from $N=18$~antennas with positions $x_j=x_A+jd/\sqrt{2}$, $y_j=y_A-jd/\sqrt{2}$ with $x_A=40$~cm, $y_A=5$~cm and
	$d=0.5$~cm are superposed;
	the center of the coordinate system is the center of the Teflon plate (see \fref{fig:whole_billiard_wave}). Because of the high resolution needed, it 		is not practical to measure the electric field on the whole Teflon plate, as it would take several years for the 
	18-antenna array. Instead, they are only measured on parts of it as indicated in \fref{fig:whole_billiard_wave}. One such part lies on the 		 		incoming beam and one on the outgoing beam; the 
	averages in~\eref{eqn:average_poynting_def} are performed over these fields. The third measured part is directly on the boundary of the Teflon plate. 		While  it is not necessary for the extraction of the GHS, the data measured here allow the direct observation of the reflection process.
	\Fref{fig:whole_billiard_wave} shows the measured real parts $\text{Re}(\psi)$ of the beams  generated according to 
	\eref{eqn:plane_wave_phase_factors}
	with $\phi_j=0$ at a frequency of $\nu=15$~GHz (corresponding to a vacuum wavelength $\lambda\approx 2$~cm). The boundary of the Teflon plate is 		shown, as well as the position of the $j=0$ antenna in the middle of the antenna array. It is clear from \fref{fig:whole_billiard_wave} that both the 		incoming and the outgoing beams have nearly plane wave fronts and travel at an angle of approximately 45~degrees to the plate boundary.	

%%%%%%%%%%%%%%%%%%%%%%%%%%%%%%%%%%%%%%%%%%%%%%%%%%%%%%%%%%%%%%%%%%%%
% Results
%%%%%%%%%%%%%%%%%%%%%%%%%%%%%%%%%%%%%%%%%%%%%%%%%%%%%%%%%%%%%%%%%%%%

\section{Results}
\label{sec:results}

\subsection{Generated \quoting{plane wave} beams}
\label{subsec:generated_beams}
	
\begin{figure}
		\includegraphics[width=1.1\textwidth]{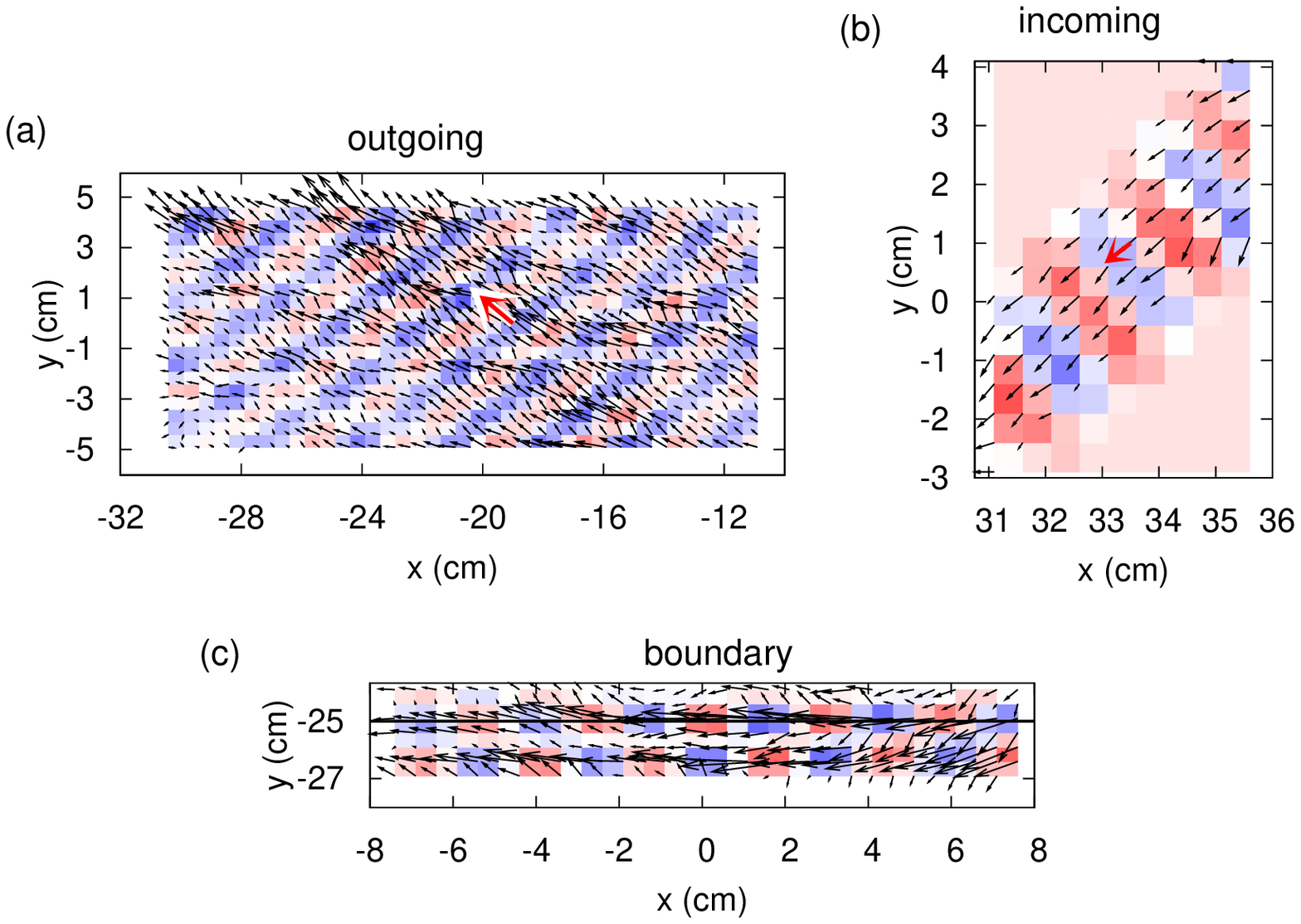}
		\caption{\label{fig:individual_waves} Real parts $\text{Re}(\psi)$  of the beams (color scale) and Poynting vectors (black arrows) in the three 		measured parts of the Teflon plate as shown in 
		\fref{fig:whole_billiard_wave}. (a) outgoing beam, (b) incoming beam, (c) beam at the plate boundary. The average Poynting vectors of the incoming 			and outgoing beams are shown as red arrows.}
	\end{figure}

	Details of the beams in the different measured parts of the Teflon plate are shown individually in
	\fref{fig:individual_waves} together with their Poynting vectors.
	The positions $\vec{r}_{\text{in}}$ and $\vec{r}_{\text{out}}$ given by
	\begin{equation}
	\label{eqn:r_in_r_out}
		\vec{r}_{\text{in}}=\pmatrix{%
				33.5\cr 1.0
			}\text{ cm}, \, \, \vec{r}_{\text{out}}=\pmatrix{%
				-19.0 \cr 0.0
			}\text{ cm}
	\end{equation}
	for the Poynting vectors in this figure.

	The spatial resolution of 5~mm is clearly sufficient to see the structure of the
	incoming and outgoing beams, which in fact approximate plane waves. 
	Their respective Poynting vectors show a well-defined propagation
	direction. The incoming angle calculated from the averaged Poynting vector is $\chi\approx 47$~degrees, the respective outgoing angle is 
	$\chi\approx 45$~degrees, which further supports the claim that, in fact, a plane wave travelling at an angle of 45~degrees has been created. The Poynting
	vectors of the beam at the boundary reveal -- perhaps surprisingly given that so little can be seen in $\text{Re}\psi$ itself -- that
	there are incoming and outgoing parts of the wave at the boundary. A part of it is reflected at the boundary, but another part penetrates outside the 		Teflon plate. The penetration depth seems to be a bit larger than the wavelength of 2~cm. 
		
	The generation of plane waves with one propagation direction thus works, at least for high $\nu$ values ($\nu\geq 10$~GHz; below that value, the wave fronts
	are less well defined and the extraction of Poynting vectors is thus not possible with high accuracy). 

\subsection{Extracted GHS}
\label{subsec:extracted_ghs}

	Two plane waves generated according to the scheme discussed in the previous section can now be superposed. The GHS of the resulting beam can be extracted 		and compared to the Artmann result. \Tref{tab:phi_j_chi} shows the different phase functions $\phi_1(j),\, \phi_2(j)$ used in this section and the 		angle of incidence $\chi$
	of the beam  according to \eref{eqn:exp_artmann3}. The choice of the $\phi(j)$ functions is of course somewhat arbitrary; here, they are
	chosen such that the $j$ dependence is simple, the difference between $\phi_1(j)$ and $\phi_2(j)$ is small (as this is the approximation in the Artmann 	result), and such that a sufficiently broad range of angles of incidence results.
	\begin{table}
		\caption{\label{tab:phi_j_chi} Phase functions $\phi_{1,2}(j)$ and the resulting incoming angles $\chi$ in degrees as extracted from the
		experimental data.}
		\begin{indented}
			\item[]\begin{tabular}{@{}l|llllllll}
				\br
				$\phi_1(j)$ & $8j/9$ & $2j/3$ & $j/2$ & $0$ & $j/10$ & $0$ & $-j/2$ & $-8j/9$ \\
				$\phi_2(j)$ & $j$ & $3j/4$ & $j/3$ & $j/4$ & $0$ & $-j/4$ & $-j/3$ & $-j$ \\
				$\chi$ (degrees) & 22.5 & 27.9 & 35.1 & 41.4 & 44.9 & 47.6 & 55.9 & 68.9 \\
				\br
			\end{tabular}
		\end{indented}
	\end{table}

\begin{figure}
		\includegraphics[width=0.8\textwidth]{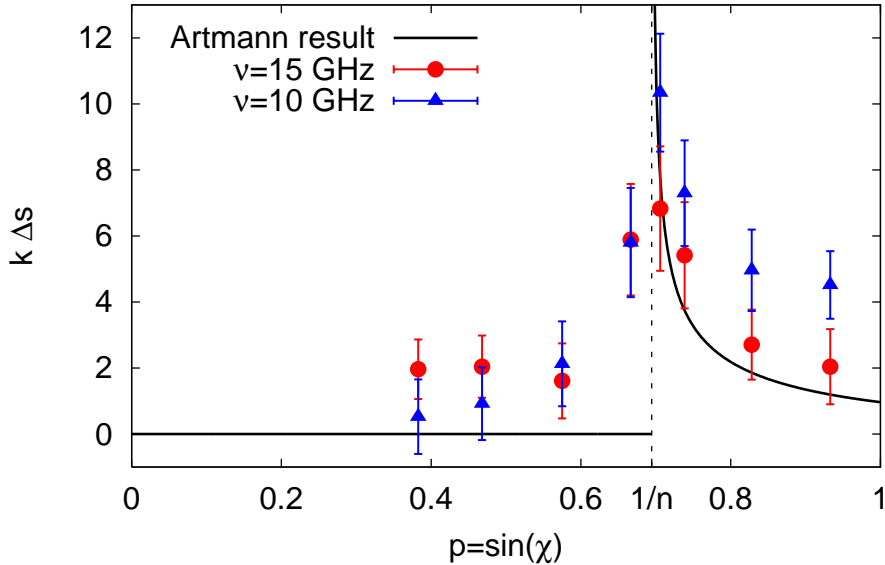}
		\caption{\label{fig:ghs_data} GHS~$k\Delta s$ as extracted from the measured data for $\nu=15$~GHz (red dots) and $\nu=10$~GHz (blue triangles). The 			black solid line is the Artmann result, the black dashed line marks the critical angle for total internal reflection for $n=1.44$.}
	\end{figure}

	The extracted GHS for the incoming angles given in \tref{tab:phi_j_chi} is shown in \fref{fig:ghs_data} for $\nu=10$~GHz ($\lambda\approx 3$~cm) and 
	$\nu=15$~GHz ($\lambda\approx 2$~cm) together with the Artmann result \eref{eqn:artmann}, which for TM 		 		polarization is given by
	\begin{equation}
	\label{eqn:artmann}
		\Delta s=\frac{1}{nk\,\sin\chi}\frac{1}{\sqrt{\sin^2\chi-1/n^2}},
	\end{equation}
	where $k=2\pi/\lambda$ is the wave number. $k\Delta s$ is thus independent of the wavelength.
 	The error bars are the errors in $\langle\vec{S}\rangle_{\text{in (out)}}$,
	which in turn, as $\langle\vec{S}\rangle_{\text{in (out)}}$ is an average value, are given by the standard deviation. The error $\delta \Delta s$ in $\Delta 		s$ is then given by error propagation; the errors are all approximately $\delta \Delta s/\Delta s\approx 20$~\% (incoming angles above the critical angle) 		and $\delta \Delta s/\Delta s\approx 50$~\% (incoming angles below critical angle). As the GHS below the critical angle is small, larger 		 		relative errors are expected.

	We see clear signatures of the GHS and the main features of the Artmann result (zero GHS below the critical angle, maximum GHS at the critical angle, 		approximately constant GHS above the critical angle, independence 
	of $k\Delta s$ of $k$) are  well captured.  At $\nu=10$~GHz, the GHS is systematically higher than the Artmann
	result predicts. This is not surprising, as deviations due to the finite width of our generated beams can be expected to become larger for smaller 		frequencies.
	
\begin{figure}
		\includegraphics[width=0.8\textwidth]{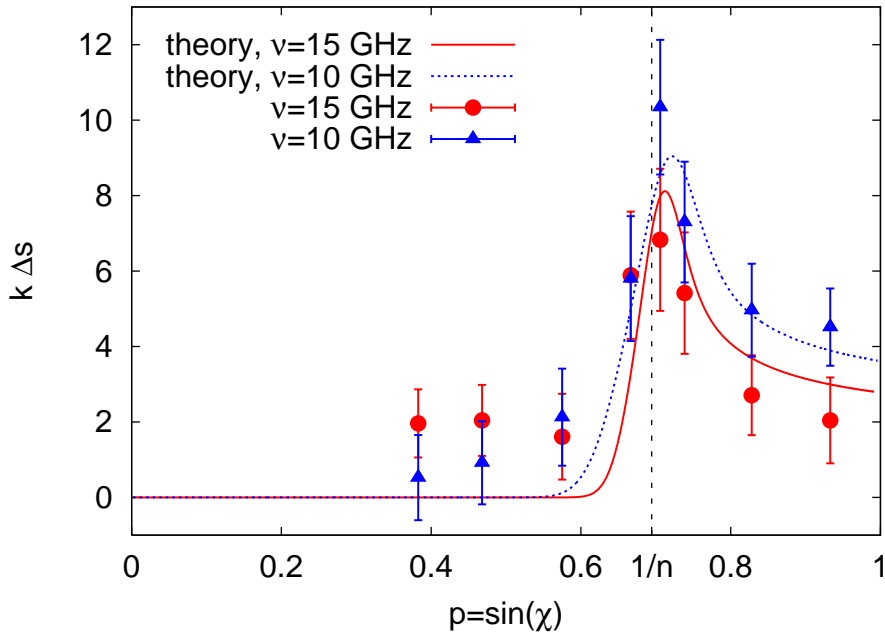}
		\caption{\label{fig:ghs_data_gaussian} GHS~$k\Delta s$ extracted from the measured data at $\nu=15$~GHz (red dots) and $\nu=10$~GHz (blue
		triangles). For comparison, the calculated GHS for a Gaussian wave packet with the corresponding frequency is shown as well (red solid and blue 		dotted curves). The black dashed line marks the critical angle for total internal reflection for $n=1.44$.}
	\end{figure}

	In fact, the agreement of the measured GHS with theoretical predictions gets even better if the finite
	width of the beams is taken into account. The individual beams are not completely plane waves, as they have a width fixed by the width of the antenna array. 		If one approximates them as Gaussian (which is justified as 
	Lai \textit{et al.} \cite{Lai1986} have shown that the precise beam profile has little influence on the
	resulting GHS) with a beam width given by the experimentally extracted value  $\sigma\approx 3$~cm, the resulting 
	GHS~$\Delta s$ can be calculated numerically as described in \cite{Unterhinninghofen2010}.
	\Fref{fig:ghs_data_gaussian} shows the
	results together with the experimental data for $\nu=10$~GHz and $\nu=15$~GHz. The deviations above the critical angle are indeed explained very well by a 		Gaussian (non-plane wave) beam profile. 

	Overall, we find convincing agreement between the GHS extracted from our experimental data and the theoretical prediction of the Artmann result. The 		deviations
	of beams with smaller frequencies from the Artmann result are due to the finite width of our generated beams, as a numerical calculation of the GHS using
	a Gaussian beam profile with a beam width corresponding to the one found in our experiments shows. 

%%%%%%%%%%%%%%%%%%%%%%%%%%%%%%%%%%%%%%%%%%%%%%%%%%%%%%%%%%%%%%%%%%%%%%%
% Conclusions
%%%%%%%%%%%%%%%%%%%%%%%%%%%%%%%%%%%%%%%%%%%%%%%%%%%%%%%%%%%%%%%%%%%%%%%

\section{Conclusions and outlook}
\label{sec:conclusions_outlook}

We presented measurements of the Goos-H{\"a}nchen shift of microwave beams at a dielectric  interface (Teflon to air). The beams were generated by superposing the spherical waves produced by 18~microwave antennas, which leads to beams with approximately plane wave fronts. The GHS is extracted from the Poynting vectors of the incoming and reflected beams, which can be found from measurements of the electric field of the beams. Our results agree well with the classical result of Artmann. 

Our experiment demonstrates that GHS measurements in dielectric microwave cavities are possible and meaningful. A similar setup using not a rectangular Teflon plate, but one with curved boundaries, could be used to study the effects of boundary curvature on the GHS. When the GHS is incorporated into ray models, it is usually assumed that the curvature effects are small, and the GHS as calculated at a planar interface can be used. Microwave measurements could give insight into the question if and when this is possible, and what deviations can be expected. Another interesting experiment would be to excite modes in a microwave cavity and, by measuring electric fields, directly observe effects of the GHS which have been predicted in optical microcavities, e.g. the formation of new modes in an elliptical cavity \cite{Unterhinninghofen2008} or the phase-space shift of periodic orbits \cite{Unterhinninghofen2010}.

%%%%%%%%%%%%%%%%%%%%%%%%%%%%%%%%%%%%%%%%%%%%%%%%%%%%%%%%%%%%%%%%%%%%%%%
% Acknowledgement
%%%%%%%%%%%%%%%%%%%%%%%%%%%%%%%%%%%%%%%%%%%%%%%%%%%%%%%%%%%%%%%%%%%%%%%

\ack
\label{sec*:acknowledgements}

Support by the DFG within the research group 760 \quoting{Scattering Systems with Complex Dynamics} and the DFG Emmy Noether Programme (M. H.) is acknowledged.

%%%%%%%%%%%%%%%%%%%%%%%%%%%%%%%%%%%%%%%%%%%%%%%%%%%%%%%%%%%%%%%%%%%%%%%
% Bibliography
%%%%%%%%%%%%%%%%%%%%%%%%%%%%%%%%%%%%%%%%%%%%%%%%%%%%%%%%%%%%%%%%%%%%%%%

\section*{References}
\label{sec*:references}

\bibliography{bibliography_paper_ghs_exp_final.bib}

%%%%%%%%%%%%%%%%%%%%%%%%%%%%%%%%%%%%%%%%%%%%%%%%%%%%%%%%%%%%%%%%%%%%%%%

\end{document}